\documentclass{aip-cp}

\usepackage[numbers]{natbib}
\usepackage{rotating}
\usepackage{graphicx}
\usepackage{amssymb} 
\usepackage{amsfonts}
\usepackage{slashed} 
\usepackage{epstopdf}
\usepackage{color}
\usepackage{caption}

\definecolor{darkred}{rgb}{1., 0., 0.4}
\begin{document}

\title{On the Stability of $\Lambda(1405)$ Matter}

\author[aff1,aff2]{J.~Hrt\'{a}nkov\'{a}\corref{cor1}} 
\author[aff3]{N.~Barnea} 
\author[aff3]{E.~Friedman}
\author[aff3]{A.~Gal}
\author[aff1]{J.~Mare\v{s}}
\author[aff1,aff2]{M.~Sch\"{a}fer}

\affil[aff1]{Nuclear Physics Institute, 250 68 \v{R}e\v{z}, Czech Republic}
\affil[aff2]{Faculty of Nuclear Sciences and Physical Engineering,\\ Czech Technical University in Prague, 115 19 Prague 1, Czech Republic}
\affil[aff3]{Racah Institute of Physics, The Hebrew University, 91904 
Jerusalem, Israel}
\corresp[cor1]{Corresponding author: hrtankova@ujf.cas.cz}

\maketitle

\begin{abstract}
A hypothesis of absolutely stable strange hadronic matter composed of $\Lambda(1405)$ baryons, here denoted $\Lambda^*$, is tested within many-body calculations performed using the Relativistic Mean-Field approach. In our calculations, we employed the $\Lambda^*\Lambda^*$ interaction compatible with the $\Lambda^*\Lambda^*$ binding energy $B_{\Lambda^*\Lambda^*}=40$~MeV given by the phenomenological energy-independent $\bar{K}N$ interaction model by Yamazaki and Akaishi (YA). We found that the binding energy per $\Lambda^*$, as well as the central density in $\Lambda^*$ many-body systems saturates for mass number $A\geq120$, leaving $\Lambda^*$ aggregates highly unstable against strong interaction decay. Moreover, we confronted the YA interaction model with kaonic atom data and found that it fails to reproduce the $K^-$ single-nucleon absorption fractions at rest from bubble chamber experiments.
\end{abstract}

\section{INTRODUCTION}
Recently, Akaishi and Yamazaki (AY) proposed that the strange matter composed of $\Lambda^*$ baryons should become increasingly bound with the number $A=-S$ of $\Lambda^*$ particles, reaching absolute stability for $A\geq8$ \cite{AY17}. This assumption is based on an energy-independent $\bar{K}N$ interaction model which identifies $\Lambda(1405)$ with the $I=0$ $\bar{K}N$ quasibound state 27 MeV below the $K^-p$ threshold \cite{YA07}. However, the implications for stable $\Lambda(1405)$ matter are based on few-body calculations with purely attractive $\Lambda^*\Lambda^*$ interactions which necessarily lead to collapse with binding energy per $\Lambda^*$ and central $\Lambda^*$ density diverging as the number of $\Lambda^*$ hyperons increases. No reliable study within the many-body calculational scheme that avoids the collapse have been performed so far. 

In this contribution, we report on our very recent calculations of $\Lambda^*$ many-body systems within the Relativistic Mean-Field (RMF) model \cite{HBFGMS18}. The meson-exchange $\Lambda^*$ potential was fitted to reproduce the binding energy of 2 $\Lambda^*$ system of 40 MeV predicted by AY \cite{AY17}. We also confronted the energy-independent $\bar{K}N$ interaction model by Yamazaki and Akaishi with kaonic atom data and $K^-$ single-nucleon absorption fractions from bubble chamber experiments.


\section{$\bar{K}N$ INTERACTION MODELS AND THEIR TESTS IN KAONIC ATOMS}
The description of low-energy meson-baryon interactions is currently provided either by chiral coupled-channel interaction models or by phenomenological energy-independent models. In Figure~\ref{fig1}, we present comparison of the real and imaginary parts of the free-space $K^-p$ (top panels) and $K^-n$ (bottom panels) amplitudes derived from the state-of-the-art chiral models and the energy-independent $\bar{K}N$ potential by YA \cite{YA07}. There is a good agreement of the $K^-p$ amplitudes at and above threshold in all models since their parameters were fitted to experimental data in this region. However, the amplitudes exhibit very different behavior below threshold. In the case of $K^-n$ amplitudes, the various models do not match each other even at and above threshold. 

The $\bar{K}N$ models have been applied in calculations of $\bar{K}$ quasibound states in few-body as well as many-body nuclear systems. The single-channel energy-independent $\bar K N$ potential by YA produces $I=0$ $\bar{K}N$ quasibound state about 27~MeV below the $K^-p$ threshold \cite{YA07}. On the other hand, the $\bar K N$ 
effective single-channel potential derived within the effective field theory approaches comes out energy dependent and the $\bar K N$ quasibound state is bound only by about 10~MeV\footnote{It is worth noting that chiral approaches with parameters fitted to all existing $K^-p$ low-energy data produce $\Lambda(1405)$ dynamically with two poles, the one with energy $\sim10$~MeV below threshold is closely related with $(\bar{K}N)_{I=0}$ quasibound state, whereas the YA model was fitted directly to the position of $\Lambda(1405)$ resonance.} \cite{HJ12}. The conjectures about stable $\Lambda^*$ matter are based on the energy-independent scenario \cite{MAY13} which predicts the $\Lambda^* \Lambda^*$ binding energy $B_{\Lambda^*\Lambda^*}$ to be $B(\bar{K}\bar{K}NN)_{I=0} - 2 B(\bar{K}N)_{I=0}=93~\textrm{MeV} - 2\times 27~\textrm{MeV}\approx40$~MeV. 

\begin{figure}[t] 
\includegraphics[width=0.45\textwidth]{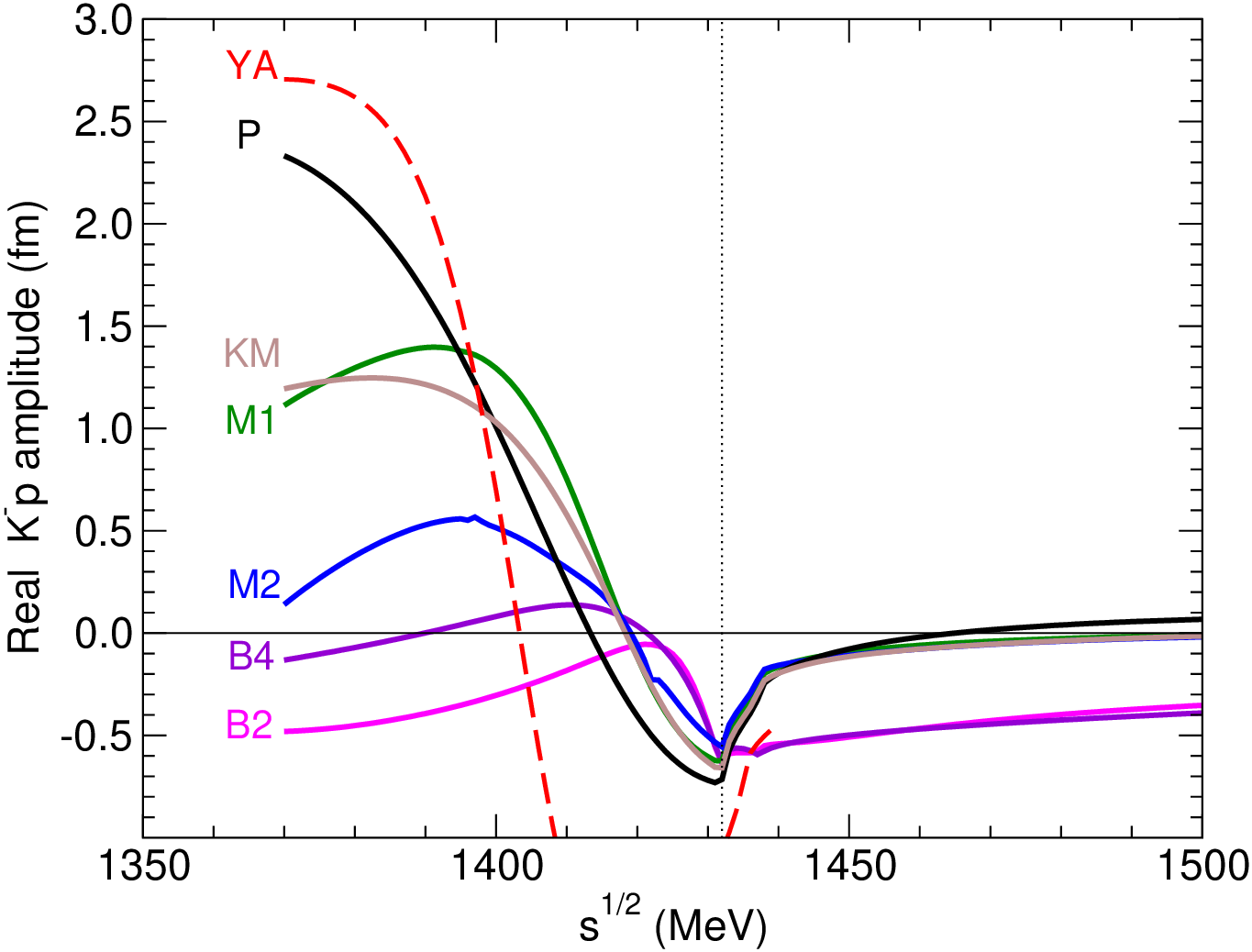} \hspace{3pt}
\includegraphics[width=0.45\textwidth]{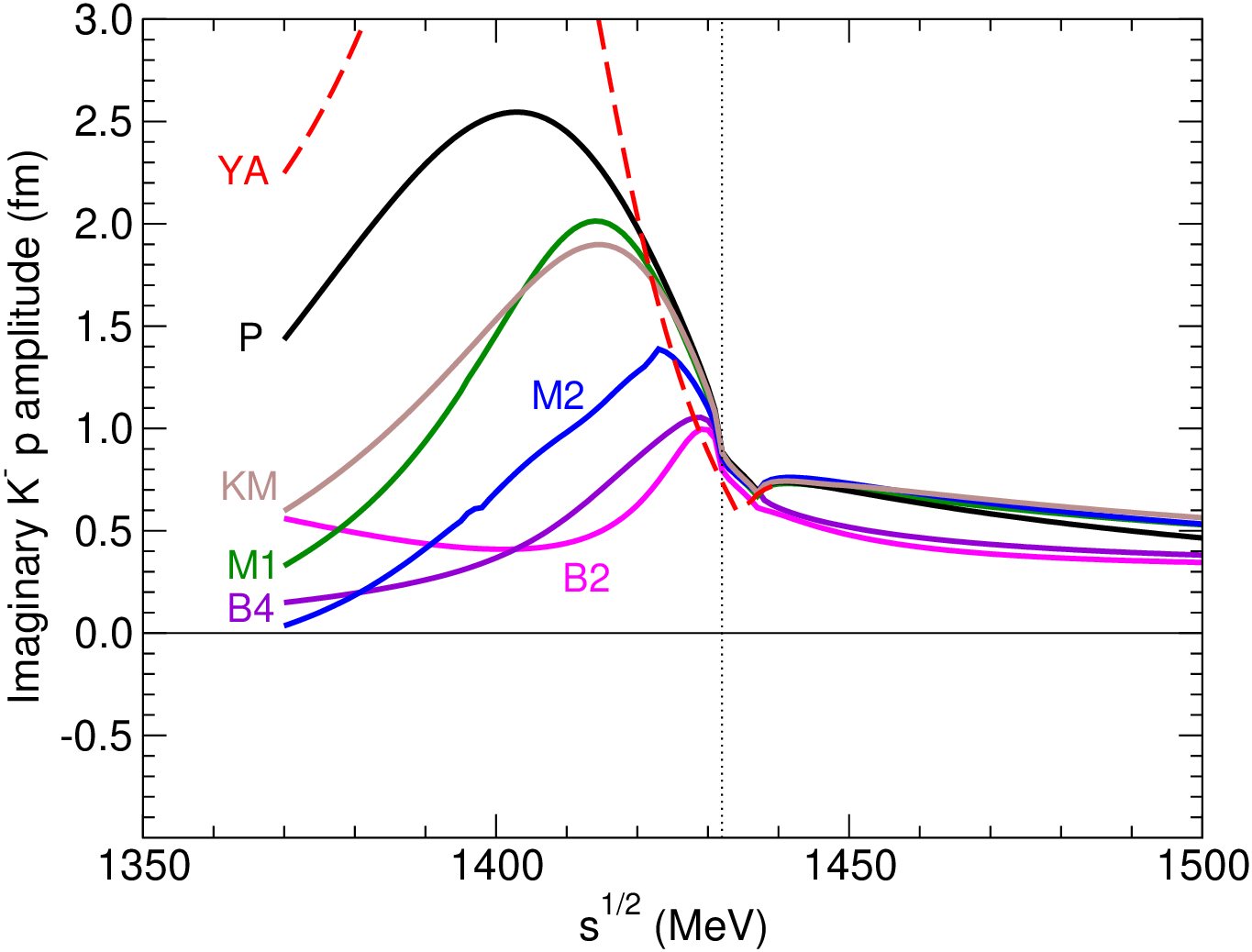} 
\end{figure}
\begin{figure}[t]
\includegraphics[width=0.45\textwidth]{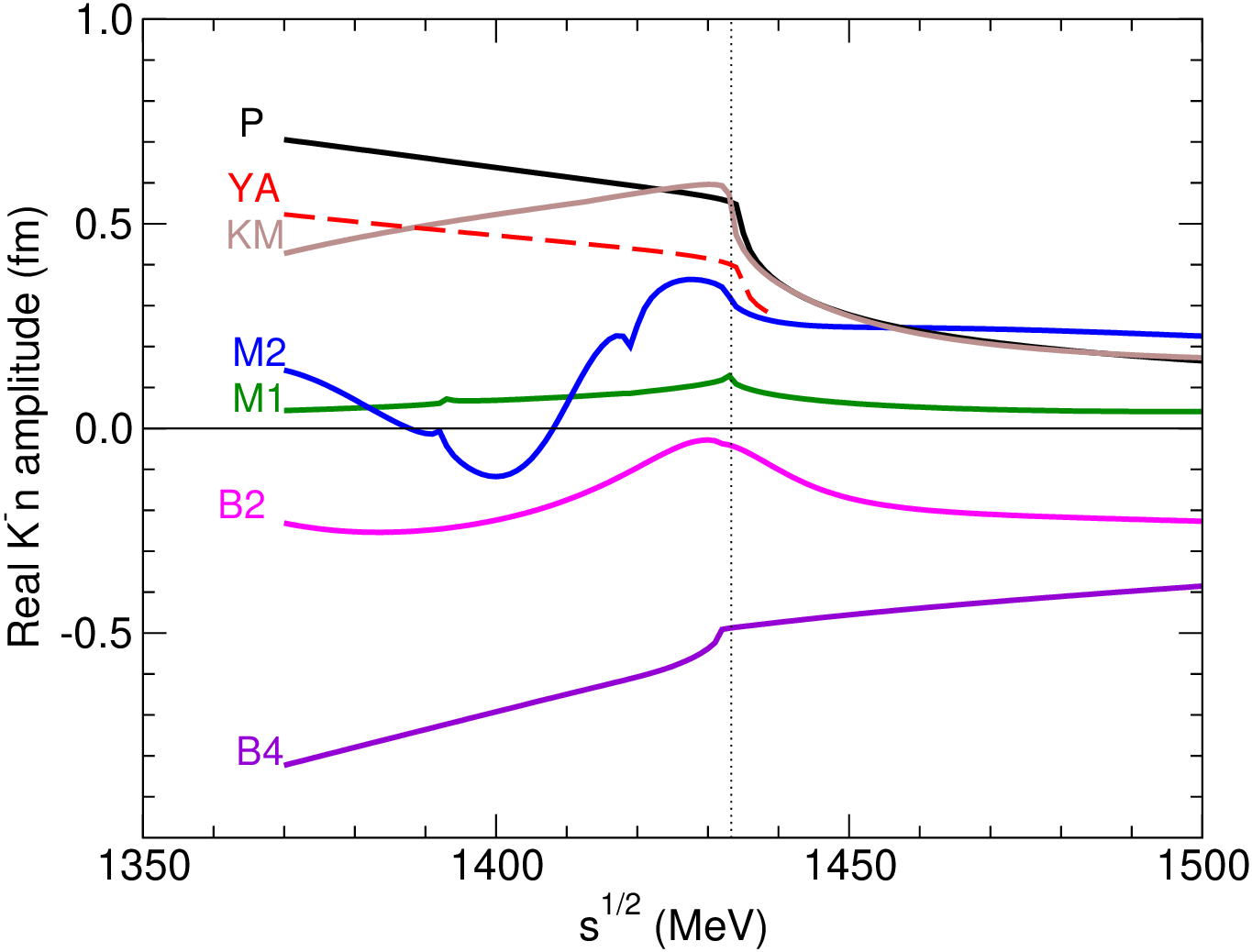} \hspace{3pt}
\includegraphics[width=0.45\textwidth]{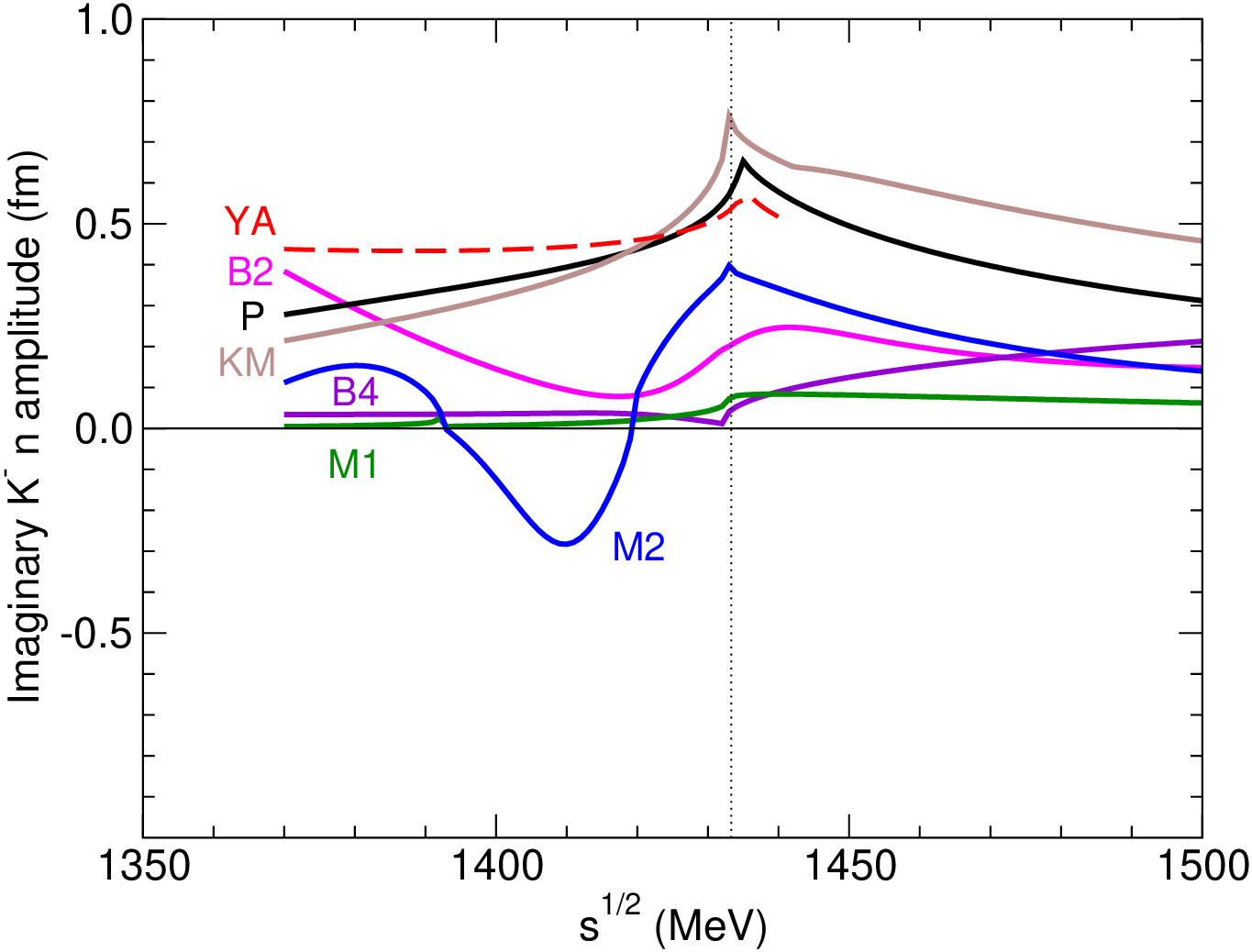} 
\caption{Real (left) and imaginary (right) parts of the free-space $K^-p$ (top) and $K^-n$ (bottom) scattering amplitudes derived from the Prague (P) \cite{pnlo}, Kyoto-Munich (KM) \cite{kmnlo}, Murcia (M1 and M2) \cite{m}, and Bonn (B2 and B4) \cite{b} chiral models and the phenomenological model by Yamazaki and Akaishi (YA) \cite{YA07}. }
\label{fig1}
\end{figure}

\begin{table}[b]
\caption{Values of $\chi^2(65)$ obtained in fits to kaonic atoms in all models considered.}
\begin{tabular}{c|ccccccc}
model & B2 & B4 &M1 & M2 & P & KM& {\color{red}YA}\\ \hline 
{\color{black}$\chi^2(65)$}  & 111  & 105 & 121 & 109 & 125 & 123 & {\color{red}150}\\
\hline
\end{tabular}
\label{tab1}
\end{table}
A sensitive test of $\bar K N$ interaction models near threshold is their ability to fit the broad data base of strong interaction energy level shifts and widths in kaonic atoms. Optical potentials based on $K^-$ single-nucleon amplitudes are generally unable to fit the kaonic atom data unless an additional phenomenological amplitude representing the interaction of $K^-$ with two or more nucleons is taken into account. This procedure was applied to several chiral $\bar K N$ amplitudes recently \cite{FG17}. In the present work we confront the YA amplitudes derived from energy-independent $\bar{K}N$ potential \cite{YA07} with kaonic atom data as well. As in the case of chiral models, the optical potential based on YA amplitudes fails to fit kaonic atom data on its own. Adding a phenomenological density-dependent $K^-$ multinucleon potential produces fits with $\chi^2$ of 150 for 65 data points. However, this $\chi^2$ is much worse than in the case of considered chiral models as shown in Table~\ref{tab1}.

\begin{figure}[t] 
\includegraphics[width=0.55\textwidth]{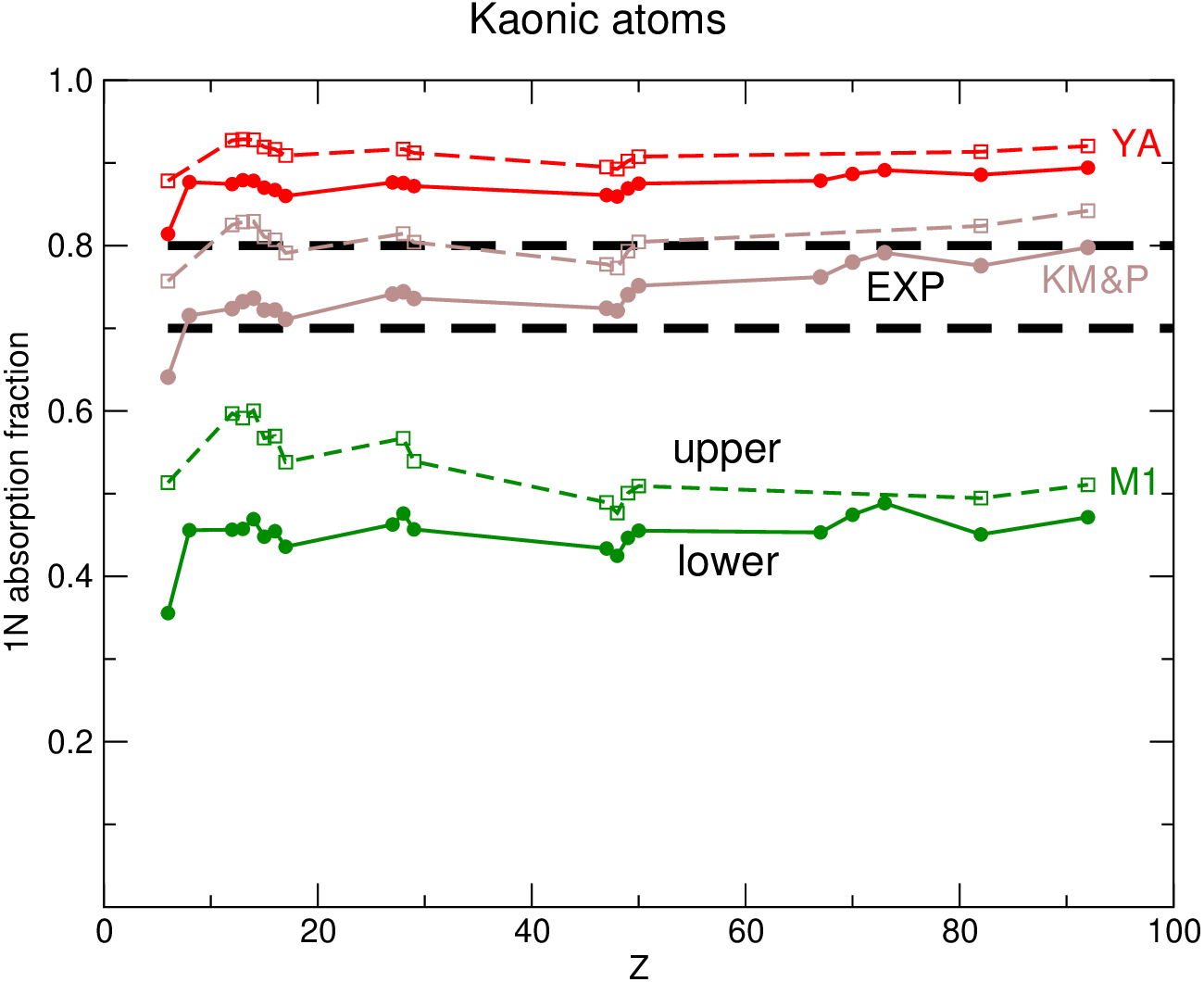}
\caption{Calculated $K^-$ single-nucleon absorption fractions in the YA model compared with the KM, P, and M1 chiral models. The range of experimentally deduced fractions is marked by two horizontal dashed lines.}
\label{fig2}
\end{figure}

Data on the $K^-$ single-nucleon absorption fractions at threshold from bubble chamber experiments \cite{DOK68,MGT71,VSW77} provide additional constraint on $\bar{K}N$ interaction models as was shown in Ref.~\cite{FG17}. In Figure~\ref{fig2}, we present calculated $K^-$ single-nucleon absorption fractions for four different models, including the YA model. The corresponding multi-nucleon amplitudes are included in these calculations as well. Solid circles correspond to the results for absorption 
from the so-called lower state and empty squares denote absorption from the upper state in each kaonic atom. As pointed out in Ref.~\cite{FG17}, not all chiral models passed this test. For example, the M1 model leads to too small ratios whereas the KM and P models provide very good agreement with experimental data (their predictions are indistinguishable from each other in the figure). The YA model yields too large ratios and thus fails to reproduce the experimental data on the $K^-$ single-nucleon absorption fractions. 

\section{$\Lambda^*$ NUCLEI}
As a next step, we explored many-body systems composed solely of $\Lambda(1405)$ baryons within the RMF approach \cite{SW86}. In our model, we considered only coupling of $\Lambda^*$ to the isocalar-scalar $\sigma$ and isoscalar-vector $\omega_{\mu}$ meson fields; the isovector-vector $\vec{\rho}$ and Coulomb fields were not taken into account since the $\Lambda^*$ is a neutral $I=0$ baryon. The Lagrangian density is of the form
\begin{equation}  
\mathcal L = \bar{\Lambda}^* \left[\,{\rm i}\gamma^\mu D_\mu-(M_{\Lambda^*}-g_{\sigma \Lambda^*} \sigma) \right] \Lambda^* + (\sigma, \omega_\mu\,\textrm{free-field terms})~,
\label{eq:Lag} 
\end{equation}
where $D_\mu=\partial_\mu+{\rm i}\,g_{\omega \Lambda^*}\,\omega_\mu$, $M_{\Lambda^*}$ is the mass of $\Lambda^*$ and $g_{i \Lambda^*}$ ($i=\sigma, \omega$) are the corresponding coupling constants. It is to be noted that we did not consider $\omega \Lambda^*$ tensor interaction since it has little effect on total binding energies of closed-shell many-body systems.
As a starting point, we employed the linear HS model \cite{HS81} for atomic nuclei to define the values of $\Lambda^*$-meson coupling constants and meson masses:
\begin{equation} 
m_{\omega}=783{\rm ~MeV},~~m_{\sigma}=520{\rm ~MeV},
~~g_{\omega \Lambda^*}=g_{\omega N}=13.80,~~g_{\sigma \Lambda^*}=g_{\sigma N}=10.47~,
\label{eq:LSH} 
\end{equation}
and performed first calculations of $\Lambda^*$ nuclei. We explored $\Lambda^*$ nuclei with closed shells and solved self-consistently the coupled system of the Klein-Gordon equations for meson fields and the Dirac equation for $\Lambda^{\ast}$.

However, in order to be consistent with the AY model we had to scale $g_{\sigma \Lambda^*}$ or $g_{\omega \Lambda^*}$ by a scaling factor $\alpha$.
We solved two-body Schr\"{o}dinger equation within the Stochastic Variational Method (SVM) \cite{VS98} for the following spin-singlet meson-exchange potentials of the Dover-Gal \cite{DG84} or Machleidt form \cite{Mach88} for the $\Lambda^*\Lambda^*$ interaction:

1) Dover-Gal
 \begin{equation} 
V_{\Lambda^*\Lambda^*}(r) = \alpha_{\omega}^2 g_{\omega \Lambda^*}^2 \,(1-\frac{3}{8}\frac{m_{\omega}^2}{M_{\Lambda^*}^2})\,
Y_{\omega}(r) -\alpha_{\sigma}^2 g_{\sigma \Lambda^*}^2 \,(1-\frac{1}{8}\frac{m_{\sigma}^2}{M_{\Lambda^*}^2})\,
Y_{\sigma}(r)~, 
\label{eq:DG} 
\end{equation} \\

2) Machleidt
\begin{equation} 
V_{\Lambda^*\Lambda^*}(r) = \alpha_{\omega}^2g_{\omega \Lambda^*}^2 \,Y_{\omega}(r) - \alpha_{\sigma}^2g_{\sigma \Lambda^*}^2 \, 
(1-\frac{1}{4}\frac{m_{\sigma}^2}{M_{\Lambda^*}^2})\, Y_{\sigma}(r)~, 
\label{eq:Mach} 
\end{equation} 
where $Y_i(r)=\exp(-m_i r)/(4\pi r)$. In the calculations we fit either the value of $\alpha_{\sigma}$  and kept $\alpha_{\omega}$ fixed to 1 or vice versa in order to get the binding energy of the $\Lambda^*\Lambda^*$ system $B_{\Lambda^*\Lambda^*}=40$~MeV. The resulting values of the scaling parameters $\alpha_{\sigma}$ and $\alpha_{\omega}$ for both types of potentials are listed in Table~\ref{tab:scaling}. We then performed RMF calculations of $\Lambda^*$ nuclei with the rescaled $\sigma$ or $\omega$ coupling constants.
\begin{table}[h]
\centering
\caption{Values of the scaling parameters $\alpha_{\sigma}$ and 
$\alpha_{\omega}$ for $\sigma$ and $\omega$ fields, respectively.}
\begin{tabular}{l|c|c}
$V_{\Lambda^* \Lambda^*}$& $\alpha_{\sigma}$ & $\alpha_{\omega}$ \\ \hline
Machleidt & 1.0913 & 0.8889 \\
Dover-Gal & 1.0332 & 0.9750 \\ \hline
\end{tabular}
\label{tab:scaling}
\end{table}

\begin{figure}[b!] 
\includegraphics[width=0.58\textwidth]{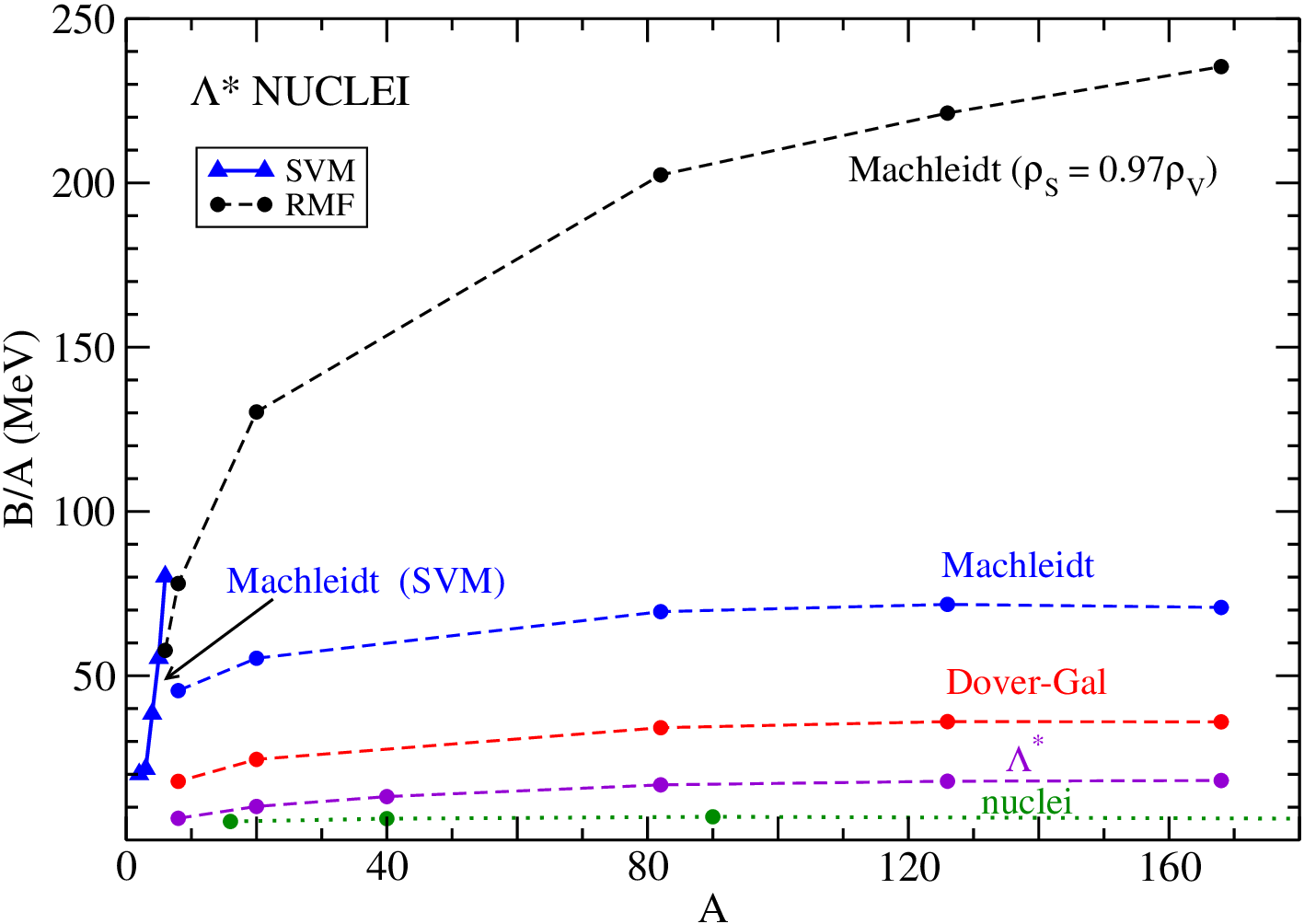}  
\caption{ Binding energy of $\Lambda^{\ast}$ nuclei per $\Lambda^{\ast}$, 
$B/A$, as a function of mass number $A$, calculated within the HS 
RMF model without scaling (denoted by `$\Lambda^*$'), with rescaled $g_{\sigma \Lambda^*}$ by $\alpha_{\sigma}$ (denoted by `Machleidt' and `Dover-Gal'), and with $\rho_{\rm s}=0.97\rho_{\rm v}$ (`Machleidt ($\rho_{\rm s}=0.97\rho_{\rm v}$)'; see text for details). The binding energy per nucleon in atomic nuclei (`nuclei') and the binding energy per $\Lambda^*$ in few-body systems calculated within the SVM by solving Schr\"{o}dinger equation for the Machleidt type potential with the rescaled $\sigma$ coupling (`Machleidt (SVM)') are shown for comparison. }
\label{fig:lstar} 
\end{figure}

The results of our calculations are illustrated in Figure~\ref{fig:lstar}. Here, the binding energy per particle, $B/A$, is plotted as a function of mass number $A$, calculated within the RMF HS model without scaling of the coupling constants (denoted by `$\Lambda^*$') and with the rescaled $\sigma$ meson coupling constant corresponding to expressions (\ref{eq:DG}) and (\ref{eq:Mach}). For comparison, the binding energy per nucleon in ordinary nuclei (denoted by `nuclei') is shown as well. The binding energy per $\Lambda^*$ saturates with the number of constituents for $A\geq120$ in all versions considered and reaches tens of MeV depending on the potential used. Calculations with rescaled $\omega$ coupling yield saturation of binding energies per $\Lambda^*$ as well with slightly larger values of $B/A$ than in the case with $\alpha_{\sigma}$. The blue triangles denote $B/A$ in few-body $\Lambda^*$ systems calculated within the SVM by solving a non-relativistic Schr\"{o}dinger equation for the Machleidt type potential~(\ref{eq:Mach}) with the scaled $\sigma$ meson coupling constant. The binding energy per $\Lambda^*$ increases rapidly in this case and does not seem to saturate with $A$ as in the RMF calculations. 

The saturation mechanism in the RMF model is driven by the Lorentz covariance which introduces two types of baryon densities --- the scalar density $\rho_{\rm s}=\bar{\psi}\psi$ associated with the attractive $\sigma$ field and the vector (baryon) density $\rho_{\rm v}=\psi^{\dagger}\psi$ associated with the repulsive $\omega$ field. The scalar density decreases in dense matter with respect to the vector density
\begin{equation}
   \rho_{\rm s} \sim \frac{M^*}{E^*} \rho_{\rm v} \quad \textrm{where} \quad \frac{M^*}{E^*} < 1~,
  \end{equation}
 and $M^*= M-g_{\sigma B} \langle \sigma \rangle$ is baryon effective mass. It means that the attraction from the scalar field is reduced in dense matter and repulsion from the vector field prevails. Saturation in RMF is thus entirely a relativistic phenomenon. In Figure~\ref{fig:lstar} we demonstrate the role of the scalar density in the saturation of $\Lambda^*$ binding energy. We performed test calculations for $\alpha_{\sigma}=1.0913$ in which we replaced the scalar density on the r.h.s of the Klein-Gordon equation for the $\sigma$ field by a density equal to 97\% of the vector density (this is the ratio of the densities in ordinary $^{16}$O). The binding energy per $\Lambda^*$ (denoted by `Machleidt $(\rho_{\rm s }=0.97\rho_{\rm v})$') is rapidly increasing in this case, similar to the SVM calculations, and does not seem to saturate within the explored mass range.

\begin{figure}[t] 
\includegraphics[width=0.58\textwidth]{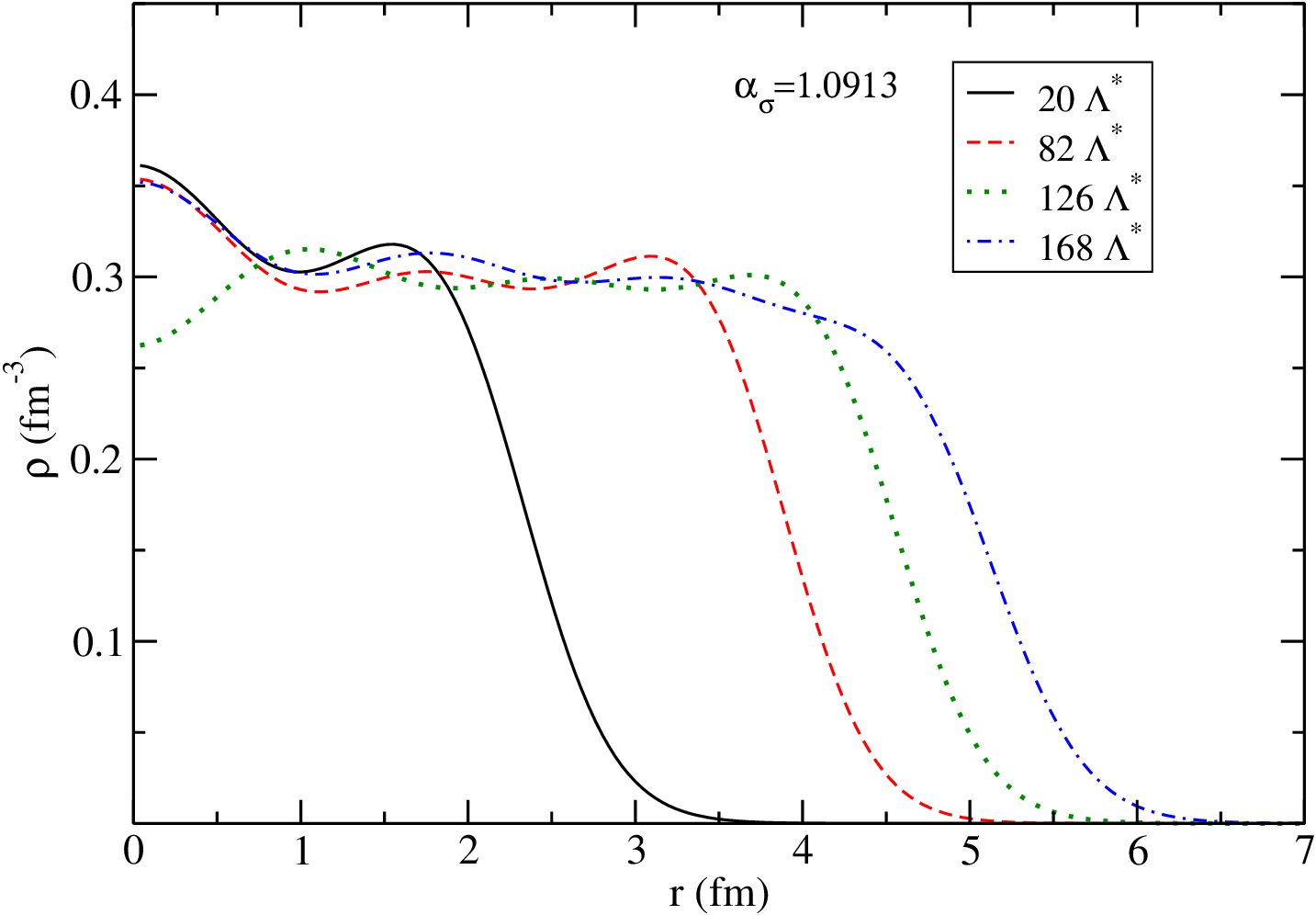}  
\caption{$\Lambda^*$ density distribution in systems composed of 20, 82, 126, and 168 $\Lambda^*$ particles, calculated within the HS model and $\alpha_{\sigma}=1.0913$.}
\label{fig:dens} 
\end{figure}
Finally, in Figure~\ref{fig:dens} we show the density distribution in $\Lambda^*$ nuclei composed of 20, 82, 126, and 168 constituents, calculated within the HS model with $\alpha_{\sigma}=1.0913$. The central density $\rho(r\approx0)$ saturates with mass number $A$ as well. It reaches values of $0.25-0.35$~fm$^{-3}$ which is about twice nuclear matter density.

\subsection{$\Lambda^*$ Decay}

Next, we performed calculations including the $\Lambda^*$ decay in order to explore how the decay width changes in the medium. We took into account the decay channel  $\Lambda^*\Lambda^* \rightarrow \Lambda \Lambda$ in the $1s$ state. The absorption was described by the imaginary part of the optical potential of the form
\begin{equation}
   \textrm{Im}V_{\rm opt}=-\frac{4 \pi}{2E_{\Lambda^*}}\frac{\sqrt{s}}{M_{\Lambda^*}} \textrm{Im}b_0~ f_{\rm supp}~ \rho~,
  \end{equation}
where $E_{\Lambda^*}=M_{\Lambda^*}+\epsilon_{\Lambda^*}$, $\sqrt{s}=\sqrt{(E_{\Lambda^*} + E_{\Lambda^*})^2-(\vec{p}_{\Lambda^*}+\vec{p}_{\Lambda^*})^2}$, Im$b_0=0.85$~fm (fitted to assumed width $\Gamma_{\Lambda^*\Lambda^*}=100$~MeV at threshold), and $f_{\rm supp}$ is the phase space suppression factor. In Table~\ref{tab:decay}, we present the single-particle energy $\epsilon_{\Lambda^*}$ and conversion width $\Gamma_{\Lambda^*\Lambda^*}$ of a $\Lambda^*$ bound in the $1s$ state in 8 and 168 $\Lambda^*$ systems. The absorption does not affect much the $\Lambda^*$ single-particle energies, they are slightly higher than in the case without the annihilation. The conversion width is suppressed in the medium, however, it still remains considerable. The $\Lambda^*\Lambda^*$ pairs in $\Lambda^*$ nuclei will inevitably decay, thus preventing to form stable baryonic matter.
  
\begin{table}[t]
\centering
\caption{1$s$ single-particle energy $\epsilon_{\Lambda^*}$ and width $\Gamma_{\Lambda^*\Lambda^*}$ (in MeV) of $\Lambda^*$ in systems composed of 8 and 168 $\Lambda^*$ baryons, calculated using the HS model with $\alpha_{\sigma}=1.0913$.}
\begin{tabular}{l|c|c}
 & 8 $\Lambda^*$ & 168 $\Lambda^*$ \\ \hline
$\epsilon_{\Lambda^*}$   &  $-133.8$  & $-202.6$\\
$\Gamma_{\Lambda^*\Lambda^*}$ & 72.1 & 99.9\\ \hline
\end{tabular}
\label{tab:decay}
\end{table}

\section{SUMMARY}

In this contribution, we explored the possibility of existence of stable $\Lambda^*$ matter based on the phenomenological energy-independent $\bar{K}N$ interaction model YA, proposed by Akaishi and Yamazaki \cite{AY17}. The YA model was confronted with kaonic atom data in the same manner as other chirally motivated $\bar{K}N$ interaction models. The optical potential based on YA amplitudes yields much worse fit to kaonic atom data than chiral models and it does not reproduce experimental values of the $K^-$ single-nucleon absorption fractions from bubble chamber experiments. We performed RMF calculations of $\Lambda^*$ nuclei with various $\Lambda^*$ interaction strengths compatible with $B_{\Lambda^* \Lambda^*}=40$~MeV of the YA model. We found that the binding energy per $\Lambda^*$ in many-body systems saturates in all cases for $A\geq 120$. The values are far below $\approx 290$~MeV, which is the energy required to reduce the $\Lambda(1405)$ mass in the medium below the mass of the lightest hyperon $\Lambda(1116)$, leaving $\Lambda^*$ aggregates unstable against strong interaction decay.


\section{ACKNOWLEDGMENTS}
J.H. and M.S. acknowledge financial support from the CTU-SGS Grant No.
SGS16/243/OHK4/3T/14. The work of N.B. is supported by the Pazy Foundation 
and by the Israel Science Foundation grant No. 1308/16.



\end{document}